\documentclass[10pt]{article}
\usepackage{amsmath}
\usepackage{graphicx}
\usepackage{color}
\usepackage{orcidlink}
\usepackage{hyperref}
\hypersetup{colorlinks=true, linkcolor=blue, citecolor=blue, urlcolor=blue}
\usepackage{subcaption}
\begin{document}
\baselineskip=16 pt

\begin{center}
{\large{\bf \textbf{Dynamics of Spin-0 (Particles-Antiparticles) in Bonnor-Melvin Cosmological Space-Time Using the Generalized Feshbach-Villars Transformation}}}
\end{center}

\vspace{0.3cm}

\begin{center}
    {\bf Abdelmalek Bouzenada\orcidlink{0000-0002-3363-980X}}\footnote{\textbf{abdelmalekbouzenada@gmail.com (corresponding author) }}\\
    \vspace{0.1cm}
    {\it Laboratory of Theoretical and Applied Physics, Echahid Cheikh Larbi Tebessi University, Algeria}\\
    \vspace{0.2cm} 
     {\bf Abdelmalek Boumali\orcidlink{0000-0002-5142-1695}}\footnote{\bf boumali.abdelmalek@gmail.com}\\
    \vspace{0.1cm}
    {\it Laboratory of Theoretical and Applied Physics, Echahid Cheikh Larbi Tebessi University, Algeria}\\
     \vspace{0.2cm}
      {\bf Faizuddin Ahmed\orcidlink{0000-0003-2196-9622}}\footnote{\bf faizuddinahmed15@gmail.com}\\
    \vspace{0.1cm}
    {\it Department of Physics, University of Science \& Technology Meghalaya, Ri-Bhoi, 793101, India}
\end{center}

\vspace{0.3cm}

\begin{abstract}
In this paper, we employ the Generalized Feshbach-Villars transformation (GFVT) to investigate the relativistic quantum dynamics of spin-0 scalar particles within the backdrop of a magnetic universe characterized by the Bonnor-Melvin cosmological space-time, which exhibits a geometrical topology resulting in an angular deficit. We derive the radial equation of the Klein-Gordon equation using this FV representation and obtain analytical solutions utilizing special functions. Our analysis demonstrates that various parameters associated with the space-time geometry exert significant influence on the eigenvalue solutions within this novel representation. This research sheds light on the intricate dynamics of particles within the theoretical framework of quantum field theory in curved space-time.
\end{abstract}

\vspace{0.1cm}

{\bf Keywords}: Spin-0: Particles-Antiparticles; Feshbach-Villars Transformation; Cosmological Constant; Quantification of Energy Spectrum; Polynomial Orthogonal

\vspace{0.1cm}

{\bf PACS Number(s):} 03.65.Pm, 03.65.Ge, 14.80.Hv, 02.30.Gp, 03.65.Vf

\section{Introduction}\label{sec1}

One of the most profound and challenging frontiers in modern physics lies at the intersection of general relativity \cite{a1,a2} and quantum mechanics \cite{a3}. This endeavor seeks to reconcile two foundational theories that elucidate the behavior of the universe at vastly different scales. In general relativity, the structure of spacetime is molded by the gravitational interactions among massive objects, while quantum mechanics governs the behavior of particles at the microscopic level \cite{a4}, including phenomena like particle-wave duality and quantum entanglement \cite{a5,t1,t2,t3,t4}. However, discrepancies emerge when attempting to apply these theories simultaneously, especially in extreme scenarios such as those near black holes \cite{a6,a6-1,a6-2,a6-3} or during the early epochs of the universe. This gives rise to enigmatic physical phenomena and poses significant questions regarding the nature of reality. The enigmas surrounding black hole singularities \cite{a7,a8,a9}, the information paradox \cite{a10}, the quantization of gravity \cite{a11,a11-1,a11-2}, and the quest for a quantum gravity theory that seamlessly integrates the principles of general relativity with those of quantum mechanics are among the most prominent challenges. The resolution of these mysteries holds the potential for paradigm-shifting discoveries that could deepen our understanding of the universe and unveil its fundamental laws.

Among the many equations that comprise the field of quantum mechanics, a few stand out as particularly important for understanding the quantum world. The Klein-Gordon equation \cite{E1}, Dirac equation \cite{E2}, Majorana equation \cite{E3}, and Duffin-Kemmer-Petiau (DKP) equation \cite{E4} hold special significance. The integration of special relativity into quantum mechanics began with the introduction of the Klein-Gordon equation, initially formulated to depict the behavior of relativistic scalar particles, laying the groundwork for quantum field theory. The Dirac equation, a foundational piece of quantum physics, united special relativity and quantum mechanics, providing a framework for explaining the behavior of relativistic spin-1/2 particles like electrons. The DKP equation, a valuable tool for characterizing systems with higher spin, is a higher-order relativistic wave equation offering insights into the behavior of exotic states and composite particles. Lastly, the Majorana equation, named after the influential physicist E. Majorana, elucidates the behavior of neutral fermions, including their antiparticles, with significant ramifications for research into dark matter and physics beyond the standard model. Taken together, these equations serve as the cornerstones of quantum physics, influencing our comprehension of the basic elements of the cosmos and providing profound insights into the nature of reality.

In the realm of quantum mechanics, the Feshbach-Villars representation holds significant importance due to its provision of a robust mathematical framework for tackling intricate theoretical challenges. Initially conceived as a solution to issues stemming from negative norm states and ghosts within quantum field theory, this representation has evolved to become widely utilized across various branches of physics, including quantum mechanics. At its core, the Feshbach-Villars representation alters the Lagrangian density through the introduction of auxiliary fields, thereby facilitating the distinction between physical and non-physical states. This distinction simplifies the treatment of quantum systems, allowing for a more systematic and comprehensible approach to developing equations of motion that accurately capture physical phenomena while circumventing the pitfalls associated with negative norm states. The application of the Feshbach-Villars representation in quantum mechanics spans diverse areas, ranging from the examination of quantum field theoretic effects in condensed matter systems to the exploration of relativistic quantum mechanics. Given its versatility and effectiveness, theoretical physicists leverage this representation as a valuable tool to deepen their understanding of the fundamental principles governing the universe and to unravel the complexities of quantum phenomena (readers are refereed to the references \cite{E5, silenko1, silenko2,silenko3, bou1, bou2, bou3, bou4, bou5,bou6, win2024,ba1}).

Moreover, despite yielding identical solutions, the FV formalism presents several advantages over directly employing the KG equation:
(i) In the KG equation, positive and negative energy solutions are intertwined, potentially leading to mathematical complexities and interpretational challenges. Contrarily, the FV formalism introduces auxiliary fields to segregate the positive and negative energy components of the wave function, resulting in a more streamlined mathematical treatment and enhanced physical intuitiveness. (ii) The FV formalism effectively eradicates negative energy solutions from the equations of motion. This is particularly advantageous as negative energy solutions within the KG equation can engender unphysical predictions. By exclusively dealing with positive energy solutions, such issues can be circumvented. Additionally, the FV formalism finds frequent application in scenarios involving electromagnetic interactions, where it aids in preserving gauge invariance and simplifying the treatment of particles with spin. (iii) Furthermore, the FV formalism offers a probabilistic interpretation and furnishes a rationale for negative energies within the quantum mechanics of relativistic particles devoid of spin. (iv) Finally, the FV formalism streamlines the handling of relativistic particles and fields, especially within the realm of quantum field theory. By segregating positive and negative energy solutions, eliminating undesirable negative energy states, and upholding Lorentz covariance, it serves as a valuable instrument for tackling a multitude of challenges in relativistic quantum physics.


The Klein-Gordon formula can be written as follows \cite{k1,k2,k3}:
\begin{equation}
\left(\frac{1}{\sqrt{-g}}\partial_{\mu}\left(\sqrt{-g}g^{\mu\nu}\partial_{\nu}\right)+M^{2}-\zeta\,R\right)\Phi(\textbf{r})=0,\label{a1}
\end{equation}
where $R$ signifies the Ricci scalar curvature, which is defined as $R=g^{\mu\nu}R_{\mu\nu}$, where $R_{\mu\nu}$ is the Ricci curvature tensor $\xi$ is a real dimensionless coupling constant, $g^{\mu\nu}$ represents the inverse metric tensor of $g_{\mu\nu}$, and $g=\det\left(g_{\mu\nu}\right)$ is its determinant. 

In this analysis, we are very much interested on a cosmological space-time generated by a magnetic field. The line-element describing this cosmological space-time is given by \cite{ML1, ML2, ML3, ML4, ML5}.
\begin{equation}
ds^{2}=g_{\mu\nu}\,dx^{\mu}\,dx^{\nu}=-dt^{2}+d\rho^{2}+\sigma^{2}\,\sin^{2}(\sqrt{2\,\Lambda}\,\rho)\,d\varphi^{2}+dz^{2},\label{a2}
\end{equation}

For the line-element (\ref{a2}), the metric tensor $g_{\mu\nu}$ and its inverse $g^{\mu\nu}$ are provided by:
\begin{equation}
g_{\mu\nu}=\left(\begin{array}{cccc}
-1 & 0 & 0 & 0\\
0 & 1 & 0 & 0\\
0 & 0 & \sigma^2\,\sin^2(\sqrt{2\,\Lambda}\,\rho) & 0\\
0 & 0 & 0 & 1
\end{array}\right),\quad 
g^{\mu\nu}=\left(\begin{array}{cccc}
-1 & 0 & 0 & 0\\
0 & 1 & 0 & 0\\
0 & 0 & \frac{1}{\sigma^2\,\sin^2(\sqrt{2\,\Lambda}\,\rho)} & 0\\
0 & 0 & 0 & 1
\end{array}\right).\label{a3}
\end{equation}

Here, $\sigma$ is a topological parameter controlling an angular deficit, and $\Lambda$ is the cosmological constant. The following relation connects these characteristics to the strength of the magnetic field along the symmetry axis:

\begin{equation}
H (\rho)=\sigma\,\sqrt{\Lambda}\,\sin\left(\sqrt{2\,\Lambda}\,\rho\right),\quad \Lambda>0.\label{a4}
\end{equation}

The determinant of the metric tensor $g_{\mu\nu}$
  for the space-time \eqref{a3} is expressed as:
\begin{equation}
\sqrt{-g}=\sigma\,\sin(\sqrt{2\,\Lambda}\,\rho).\label{a5}
\end{equation}


Our main motivation is to investigate the relativistic quantum motion of spin-0 scalar particles within the FV representation in the background of curved space-time (\ref{a2}). The complex dynamical system is examined in curved space-time generated by a magnetic field aligned along the symmetry axis under the influence of geometrical topology, specifically the Bonnor-Melvin-Lambda (BML) solution. The magnetic field holds wide-ranging applications in the context of astrophysical objects, including pulsars and quasars. We aim to analyze the quantum motions of scalar particles in this particular BML magnetic universe background within FV-representation and discuss the influence of the cosmological constant and the topological parameter on the eigenvalue solutions, providing a comprehensive understanding of the quantum system under investigation.      


\section{Quantum dynamics of spin-0 particle in Bonnor-Melvin Cosmological space-time: Generalized Feshbach-Villars Transformation (GFVT) } \label{sec3}

The GFVT is used for the characterization of both massless and massive particles. An equivalent alteration was previously presented in Ref. \cite{silenko1}. We shall follow the methods outlined in the preceding references to obtain the Feshbach-Villars (FV) form of the Klein-Gordon wave equation on curved manifolds. The following sources provide the components of the wave function $\Phi$:
\begin{equation}
i\,\tilde{\mathcal{O}}\,\psi=\mathcal{P}\,\left(\phi_{1}-\phi_{2}\right), \quad\quad\psi=\phi_{1}+\phi_{2},\label{a6}
\end{equation}
Whereas $\tilde{\mathcal{O}}=\Big(\frac{\partial}{\partial t}+\mathcal{Y}\Big)$ is defined as an arbitrary nonzero real parameter \cite{silenko3} and 
\begin{equation}
\mathcal{Y}=\frac{1}{2g^{00}\sqrt{-g}}\left\{ \partial_{i}\,,\,\sqrt{-g}g^{0i}\right\} .\label{a7}
\end{equation}

The curly braces in Equation \eqref{a7} represent the anti-commutator. After applying the transformations described earlier, the Hamiltonian can be presented in a more concise manner as \cite{silenko3}:
\begin{equation}
\mathcal{H}= \tau_{z}\,\left(\frac{\mathcal{P}^{2}+\mathcal{T}}{2\,\mathcal{P}}\right)+i\,\tau_{y}\,\left(\frac{-\mathcal{P}^{2}+\mathcal{T}}{2\,\mathcal{P}}\right)-i\,\mathcal{Y},
\label{a8}
\end{equation}

In this context, we introduce the definition \cite{silenko3}:

\begin{equation}
  \mathcal{T}= \frac{1}{g^{00}\,\sqrt{-g}}\,\partial_{i}\Big(\sqrt{-g}g^{ij}\,\partial_{j}\Big)+\frac{M^{2}-\zeta\,R}{g^{00}}-\mathcal{Y}^{2}, \quad\quad (i,j=1,2,3).\label{a9}  
\end{equation}

Here, the parameter $\mathcal{T}$ is interpreted as the squared kinetic energy. In Minkowski space, it equals $p^2 + m^2$.

To get over this restriction, we make use of the Generalized Feshbach-Villars Transformation (GFVT), first presented by Silenko \cite{
 silenko1}. This transformation uses a global parameter $\mathcal{P}$ that can accommodate both massless and heavy particles. The local flatness of the space-time is confirmed by the metric (\ref{a2}), as vanishing coupling terms result from the null Riemann tensor. $\mathcal{O}^{\prime}\psi=\mathcal{P}\left(\phi_{1}-\phi_{2}\right)$ and $\psi=\phi_{1}+\phi_{2}$ are obtained if $\mathcal{Y}\neq0$. Here, $\mathcal{O}^{\prime}=\frac{\partial}{\partial t}+\mathcal{Y^{\prime}}$ is obtained using \cite{silenko3}:

 \begin{equation}
     \mathcal{Y}^{\prime}=\frac{1}{2}\left\{ \partial_{i}\,,\,\frac{g^{0i}}{g^{00}}\right\}
     \label{a10}
 \end{equation}
The technique can be applied to spinning cosmic strings by first creating the equations of motion using the GFVT, then Using the GFVT to build the equations of motion and then solving them to get wave functions and energy spectra is how the technique is extended to spinning cosmic strings. In this case, determining the Hamiltonian requires a two-part formulation of the Klein-Gordon-type fields, which requires extra definitions for quantities given in earlier equations \cite{silenko3}.

\begin{equation}
\mathcal{H}^{\prime}=\tau_{z}\,\left(\frac{\mathcal{P}^{2}+\mathcal{T^{\prime}}}{2\,\mathcal{P}}\right)+i\,\tau_{y}\,\left(\frac{-\mathcal{P}^{2}+\mathcal{T^{\prime}}}{2\,\mathcal{N}}\right)-i\,\mathcal{Y}^{\prime},\label{a11}
\end{equation}
with \cite{silenko3}
\begin{align}
\mathcal{T}^{\prime} & =\partial_{i}\frac{G^{ij}}{g^{00}}\partial_{j}+\frac{M^{2}-\zeta\,R}{g^{00}}+\frac{1}{\mathcal{F}}\nabla_{i}\left(\sqrt{-g}\,G^{ij}\right)\nabla_{j}\left(\frac{1}{\mathcal{F}}\right)+\sqrt{\frac{\sqrt{-g}}{g^{00}}}\,G^{ij}\,\nabla_{i}\,\nabla_{j}\left(\frac{1}{\mathcal{F}}\right)+\frac{1}{4\,\mathcal{F}^{4}}\left[\nabla_{i}\,\mathcal{U}^{i}\right]^{2}\nonumber \\
 & -\frac{1}{2\,\mathcal{F}^{2}}\nabla_{i}\left(\frac{g^{0i}}{g^{00}}\right)\nabla_{j}\,\mathcal{U}^{j}-\frac{g^{0i}}{2\,g^{00}\,\mathcal{F}^{2}}\,\nabla_{i}\,\nabla_{j}\,\mathcal{U}^{j},\label{a12}
\end{align}
where \cite{silenko3}
\begin{equation}
G^{ij}=g^{ij}-\frac{g^{0i}g^{0j}}{g^{00}},\;\;\mathcal{F}=\sqrt{g^{00}\sqrt{-g}},\;\;\mathcal{U}^{i}=\sqrt{-g}g^{0i}.\label{a13}
\end{equation}
A non-unitary transformation $\Psi\equiv\Phi^{\prime}=\mathcal{F}\Phi$ is applied to the field $\Psi$, allowing a pseudo-Hermitian Hamiltonian to be derived. Whereas
\begin{equation}
\mathcal{H}_{GFVT}^{\prime}=\tau_{z}\left(\mathcal{H}_{GFVT}^{\prime}\right)^{\dagger}\tau_{z},\quad\quad \mathcal{H}{GFVT}^{\prime}=\mathcal{F}\mathcal{H}_{GFVT}^{\prime}\mathcal{F}^{-1}.
\end{equation}
Further algebraic operations yield the following expression for Eq. \eqref{a2}:
\begin{equation}
\mathcal{T}^{\prime}=-\frac{1}{\mathcal{F}}\left[\left(\frac{\partial}{\partial r}\right)\left(\sqrt{-g}\frac{\partial}{\partial r}\right)+\sqrt{-g}\left(\frac{1}{\sigma^{2}\,\sin^{2}\left(\rho\sqrt{2\Lambda}\right)}\right)\frac{\partial^{2}}{\partial\varphi^{2}}+\sqrt{-g}\frac{\partial^{2}}{\partial z^{2}}\right]\frac{1}{\mathcal{F}}+\frac{M^{2}}{g^{00}},\label{a14}
\end{equation}
where $R=\frac{4\,\zeta}{\Lambda}$. We can now begin to solve the subsequent equation.
\begin{equation}
\mathcal{H}_{GFVT}^{\prime}\Psi^{\prime}(\boldsymbol{x})=i\,\frac{d}{dt}\Psi^{\prime}(\boldsymbol{x}),\; \text{with}\;\boldsymbol{x}\equiv(t,\rho,\varphi,z),\label{a15}
\end{equation}
where the two equations are used to calculate $\mathcal{H}_{GFVT}^{\prime}$. As well as (\ref{a12}).

Therefore, let's consider the following ansatz for the wave function to solve this eigenvalue problem:
\begin{equation}
\Psi(\boldsymbol{x})=\mathcal{F}\Phi\left(\boldsymbol{x}\right)=\mathcal{F}\left(\begin{array}{c}
\phi_{1}\left(\boldsymbol{r}\right)\\
\phi_{2}\left(\boldsymbol{r}\right)
\end{array}\right)e^{-\left(iEt-\ell\varphi-k_{z} z\right)},\label{a16}
\end{equation}
The angular momentum quantum number in this case is $\ell=0,\pm1,\pm2,\pm3,\ldots$, and $k_{z}\in \left[ -\infty,\infty \right] $. We obtain the differential equations by substituting Equation \eqref{a14} into Equation \eqref{a11}.

\begin{equation} 
\begin{aligned}\mathcal{P}^{2}\mathcal{F}\left(\phi_{1}-\phi_{2}\right)+\mathcal{T}^{\prime}\mathcal{F}\left(\phi_{1}-\phi_{2}\right) & =2\mathcal{P}E\mathcal{F}\phi_{1},\\
\mathcal{P}^{2}\mathcal{F}\left(\phi_{1}-\phi_{2}\right)-\mathcal{T}^{\prime}\mathcal{F}\left(\phi_{1}+\phi_{2}\right) & =2\mathcal{P}E\mathcal{F}\phi_{2}.
\end{aligned}
\label{a17}
\end{equation}
By adding and subtracting these equations and then simplifying, we arrive at sets of coupled equations for $\phi_{1}$ and $\phi_{2}$.
\begin{align}
&\mathcal{P}\mathcal{F}\left(\phi_{1}-\phi_{2}\right) =E\mathcal{F}\left(\phi_{1}+\phi_{2}\right),\label{eq:34}\\
-&\mathcal{T}^{\prime}\mathcal{F}\left(\phi_{1}+\phi_{2}\right)=\mathcal{P}E\mathcal{F}\left(\phi_{1}-\phi_{2}\right),\label{a18}
\end{align}
Utilizing the obtained result, we have:
\begin{equation}
\phi_{1}=  \frac{1}{\mathcal{P}}\left(E\right)\phi_{2}
.\label{a19}
\end{equation}
After algebraic manipulations, we obtain the following second-order differential equation for the radial function $\psi(r)$:
\begin{equation}
\varphi_{1}''+\frac{\sqrt{2\,\Lambda}}{\tan(\sqrt{2\,\Lambda}\,\rho)}\,\varphi_{1}'+\Bigg[\beta^{2}-\frac{\kappa^{2}}{\sin^{2}(\sqrt{2\,\Lambda}\,\rho)}\Bigg]\,\varphi_{1}=0,\label{a20}
\end{equation}
where 
\begin{equation}
   \beta^{2}=E^{2}-M^{2}-k_{z}^{2}-\frac{4\xi}{\Lambda}\, \quad\quad\kappa^{2}=\frac{\ell^{2}}{\sigma^{2}} \label{a21}
\end{equation}
We use the change of variables method in this specific calculus application.
\begin{equation}
    s=cos(\sqrt{2\,\Lambda}\,\rho)
    \label{a22}
\end{equation}
After performing mathematical calculations, we derive the second-order differential equation. 
\begin{equation}
    \left(1-s^{2}\right)\varphi_{1}''\left(s\right)-2s\varphi_{1}'\left(s\right)+\left(\frac{\beta^{2}}{2\,\Lambda}-\frac{\left(\frac{\kappa^{2}}{2\,\Lambda}\right)}{1-s^{2}}\right)\varphi_{1}\left(s\right)=0
    \label{a23}
\end{equation}
The exact solution of this differential equation emerges in the form of Legendre polynomials. Legendre polynomials are a set of orthogonal polynomials \cite{math1,math2,math3,math4}
\begin{equation}
    \left(1-s^{2}\right)\varphi_{1}''\left(s\right)-2s\varphi_{1}'\left(s\right)+\left(\xi\left(\xi+1\right)-\frac{\delta^{2}}{1-s^{2}}\right)\varphi_{1}\left(s\right)=0
    \label{a24}
\end{equation}
where
\begin{equation}
    \xi\left(\xi+1\right)=\frac{\beta^{2}}{2\,\Lambda}	\quad\quad \to \quad\quad \xi=\frac{-\sqrt{\Lambda}+\sqrt{2\beta^{2}+\Lambda}}{2\sqrt{\Lambda}}
    \label{a25}
\end{equation}
and
\begin{equation}
    \delta^{2}=\frac{\kappa^{2}}{2\,\Lambda} \quad\quad \to \quad\quad\delta=\frac{\kappa}{\sqrt{2\,\Lambda}}
    \label{a26}
\end{equation}
The exact solution of this differential equation is given by the Legendre Polynome, The Legendre $P$ and Legendre $Q$ polynomials provide the precise solution to this differential equation, enabling us to represent the solution as a combination of these functions.  They are essential instruments for solving a wide range of differential equations that arise in several scientific areas because of their features, which include orthogonality and recursion linkages. We can forecast the dynamics and characteristics of systems governed by differential equations and obtain insights into their behavior by utilizing the solutions offered by Legendre polynomials \cite{math1,math2,math3,math4}.
\begin{equation}
    \varphi_{1}\left(s\right)=\textit{\ensuremath{\mathcal{C}_{1}}}\mathrm{P}_{\xi}^{\delta}\!\left(s\right)+\textit{\ensuremath{\mathcal{C}_{2}}}Q_{\xi}^{\delta}\!\left(s\right)
    \label{a27}
\end{equation}
We now study the relationships of the Legendre polynomials $P_n(x)$ and $Q_n(x)$ with the hypergeometric function ($_{p}F_{q} (x) $) via two different order relations ((\ref{a28}),(\ref{a29})). These interactions clarify the complex relationships that exist between these mathematical objects, providing insight into their characteristics and actions in many contexts. By proving these connections, we advance our knowledge of the fundamental ideas behind Legendre polynomials and their connection to hypergeometric functions, opening up new avenues for our comprehension of a range of mathematical phenomena\cite{math4}.
\begin{equation}
    \mathrm{P}_{a}^{b}\left(s\right)=\frac{(s+1)^{\frac{b}{2}}}{(s-1)^{\frac{b}{2}}\Gamma(1-b)}\mathrm{\,_{2}F_{1}}\left(-a,a+1,1-b,\frac{1-s}{2}\right)
    \label{a28}
\end{equation}
and
\begin{equation}
   Q_{a}^{b}\left(s\right)=\frac{\mathrm{e}^{\mathrm{i}b\pi}\sqrt{\pi}\,\left(s+1\right)^{\frac{b}{2}}\left(s-1\right)^{\frac{b}{2}}\Gamma\left(a+b+1\right)}{\left(2s\right)^{a+b+1}\Gamma\left(\frac{3}{2}+a\right)2^{a}}\mathrm{\,_{2}F_{1}}\left(1+\frac{a+b}{2},\frac{a+b+1}{2},\frac{3}{2}+a,\frac{1}{s^{2}}\right)
    \label{a29}
\end{equation}
the hypergeometric polynomial $  {}_2F_1(a,b;c;z)$ 
  can be written as \cite{math5}:
\begin{equation}
    {}_2F_1(a,b;c;z) = \sum_{n=0}^{\infty} \frac{(a)_n (b)_n}{(c)_n} \frac{z^n}{n!}
\end{equation}
where $(a)_n$ denotes the Pochhammer symbol or rising factorial, defined as \cite{math6}: 
\begin{equation}
 (a)_n = a(a+1)(a+2) \ldots (a+n-1) = a(a+1)(a+2) \ldots (a+n-1).
\end{equation}

In quantum physics, the ensuing physical conditions of the wave function can be stated by computing the limits of two polynomials at $(s=0)$ and $ (s\to \infty$). Determining the behavior and characteristics of quantum systems relies heavily on this procedure. We can write: 
\begin{equation}
    \varphi_{1}\left(s\right)\simeq\textit{\ensuremath{\mathcal{C}_{1}}}\mathrm{P}_{\xi}^{\delta}\!\left(s\right)
    \label{a30}
\end{equation}

In the context of Feshbach-Villars formalism, the determination of the total wave function of the system is :

\begin{equation}
    \varphi_{1}\left(s\right)=\textit{\ensuremath{\mathcal{N}_{1}}}\frac{(s+1)^{\frac{\kappa}{2\sqrt{2\Lambda}}}}{(-1+s)^{\frac{\kappa}{2\sqrt{2\Lambda}}}\Gamma\left(1-\frac{\kappa}{\sqrt{\Lambda}}\right)}\mathrm{\,_{2}F_{1}}\left(\frac{\sqrt{\Lambda}+\sqrt{2\beta^{2}+\Lambda}}{2\sqrt{\Lambda}},-\frac{-\sqrt{\Lambda}+\sqrt{2\beta^{2}+\Lambda}}{2\sqrt{\Lambda}},-\frac{\sqrt{2}\,\kappa-2\sqrt{\Lambda}}{2\sqrt{\Lambda}},\frac{1-s}{2}\right)
    \label{a31}
\end{equation}
The total wave function is expressed by the given relation, where $\mathcal{N}$ ensures normalization conditions.
\begin{align}
\psi_{total}\left(s\right)= & \textit{\ensuremath{\mathcal{N}\left(\begin{array}{c}
\frac{E}{\mathcal{P}}+1\\
1-\frac{E}{\mathcal{P}}
\end{array}\right)}}\frac{\left(s+1\right){}^{\frac{\kappa}{2\sqrt{2\Lambda}}}}{\left(-1+s\right){}^{\frac{\kappa}{2\sqrt{2\Lambda}}}\Gamma\left(1-\frac{\kappa}{\sqrt{\Lambda}}\right)} G_{n}(s)
\label{a32}
\end{align}
where $G_{n}(s)$ is given by
\begin{equation}
G_{n}\left(s\right)=\mathrm{\,_{2}F_{1}}\left(\frac{\sqrt{\Lambda}+\sqrt{2\beta^{2}+\Lambda}}{2\sqrt{\Lambda}},-\frac{-\sqrt{\Lambda}+\sqrt{2\beta^{2}+\Lambda}}{2\sqrt{\Lambda}},1-\frac{\kappa}{\sqrt{2\Lambda}},\frac{1-s}{2}\right) \label{a33}
\end{equation}
This representation expresses the hypergeometric function $_{2}F_{1}(-n, b; c; s)$ as a series expansion involving the Pochhammer symbol and binomial coefficients.
\begin{equation}
 _{2}F_{1}(-n, b; c; s) = \sum_{j=0}^{n} \frac{(-n)_j (b)_j}{(c)_j j!} s^j = \sum_{j=0}^{n} (-1)^j \binom{n}{j} \frac{(b)_j}{(c)_j} s^j.
\end{equation}
The condition of quantification in this wave function hypergeometric polynomial, as given by
\begin{equation}
\frac{\sqrt{\Lambda}+\sqrt{2\beta^{2}+\Lambda}}{2\sqrt{\Lambda}}=-n
    \label{a34}
\end{equation}

\begin{center}  
\begin{figure}
\begin{centering}
\subfloat[ ]{\centering{}\includegraphics[scale=0.6]{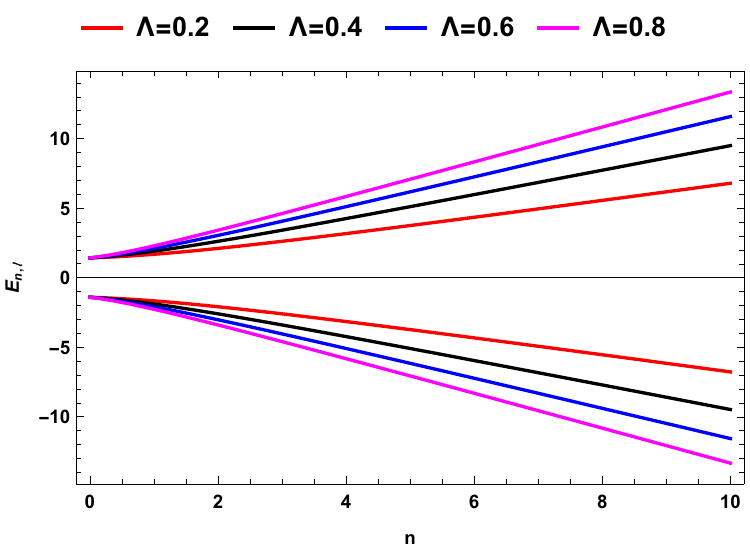}}\quad\quad
\subfloat[ ]{\centering{}\includegraphics[scale=0.6]{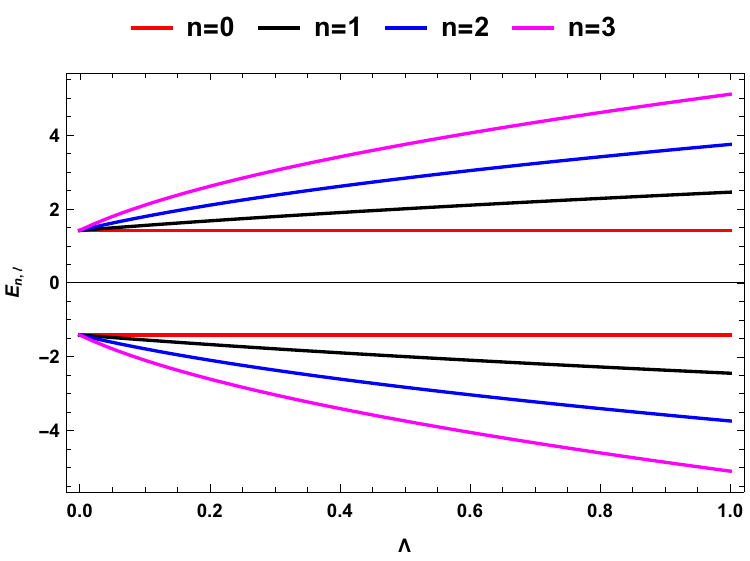}}
\par\end{centering}
\centering{}\caption{ Energy of Free Particles anti-Particles in equation (\ref{35}),where $k_{z}=M=1$ and $\xi=0$.}
\label{fig1}
\end{figure}
\par\end{center}
the energy of the free Klein-Gordon equation, examined in the Bonnor-Melvin Cosmological Space-Time using the Feshbach-Fillars Method, offers significant new understandings into the dynamics of quantum fields in curved spacetimes.
\begin{equation}
    E_{n}=\pm\sqrt{2\Lambda n\left(n+1\right)+m^{2}+k_{z}^{2}+\frac{4\xi}{\Lambda}}
    \label{35}
\end{equation}

\begin{center}  
\begin{figure}
\begin{centering}
\includegraphics[scale=0.6]{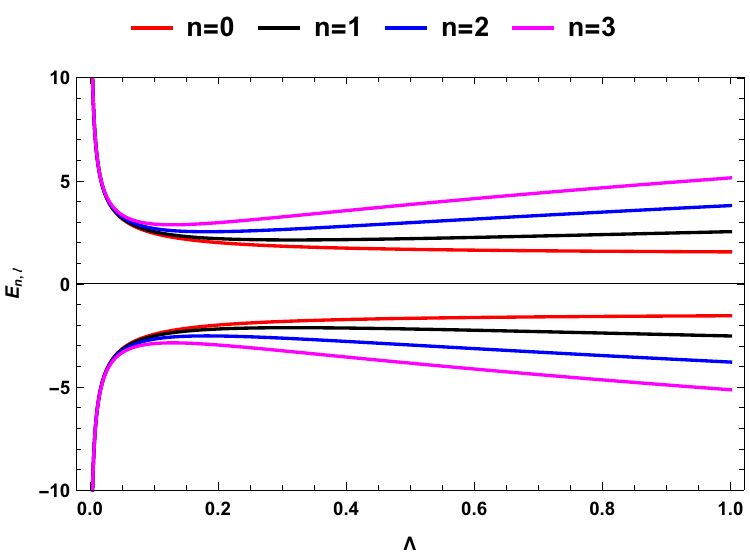}
\par\end{centering}
\centering{}\caption{ Energy of Free Particles anti-Particles in equation (\ref{35}),where $k_{z}=M=1$ and $\xi=0.1$.}
\label{fig1}
\end{figure}
\par\end{center}
The dynamics of spin-0 particles in Bonnor-Melvin cosmological spacetime with Feshbach-Villars representation is a complex topic with intriguing implications in quantum field theory and cosmology:
\begin{itemize}
\item Using the Feshbach-Villars model, the dynamics of spin-0 particles in Bonnor-Melvin cosmic spacetime have interesting theoretical physics consequences. This configuration provides a special framework for investigating particle behavior in curved spacetime.
\item A discrete spectrum of solutions, each corresponding to a different energy state of the spin-0 particles in Bonnor-Melvin spacetime, is obtained by quantifying the free equation with a quantum number $n$ (see Eq. (\ref{35})). Through the solutions of the quantization equation, the behavior of these particles may be examined, providing information about how they propagate through the cosmic background.
\item The investigation of spin-0 particles using the Feshbach-Villars representation in Bonnor-Melvin spacetime advances current efforts to integrate quantum field theory with gravitational phenomena. These kinds of projects could clarify unanswered puzzles in theoretical physics, like what dark matter is, how particles behave around black holes, and what the universe's ultimate structure is.
\item The influence of the cosmological constant in this system is profound and pervasive (see Fig.(\ref{fig1})), echoing across every level of energy within the intricate tapestry of Bonnor-Melvin cosmological spacetime. Its presence infuses the very fabric of reality, shaping the dynamics of spin-0 particles in profound ways that reverberate throughout the cosmic expanse. From the lowest energy states to the highest, the cosmological constant casts its influence, orchestrating the dance of particles with a subtle yet unmistakable hand. Its effects manifest in the quantized solutions of the system, imprinting upon them the signature of its presence and shaping the behavior of particles as they traverse the cosmic landscape. Indeed, the cosmological constant emerges as a fundamental parameter in our quest to understand the dynamics of spin-0 particles in this complex and fascinating environment, underscoring the intricate interplay between cosmic geometry and particle physics.
\item A solution to Einstein's field equations, the Bonnor-Melvin cosmological spacetime questions traditional ideas about space geometry and provides new perspectives on energy quantification in curved spacetime. This model, which shows a rotating cylindrical mass in an electromagnetic field, begs questions about the nature of energy in physics and inspires investigation into the geometry of the cosmos. Comprehending the behavior of energy in such curved spacetime contexts is essential to clarifying basic scientific phenomena, expanding our knowledge of the universe, and discovering the underlying laws guiding physical processes.
\end{itemize}

\section{Conclusion}

Our exploration delves deep into the dynamics of spin-0 particles within the complex fabric of Bonnor-Melvin cosmological spacetime, supported by the elegant Feshbach-Villars representation, revealing a captivating convergence of quantum field theory and cosmology. This multidimensional inquiry transcends mere theoretical abstraction, offering a profound journey into the essence of reality itself. With the Feshbach-Villars model as our guiding framework, we embark on a transformative odyssey through the depths of Bonnor-Melvin's cosmic expanse. Here, amidst the curvature of spacetime and the gravitational symphony it orchestrates, we uncover a profound understanding of particle behavior within this cosmic crucible, akin to navigating the intricate pathways of a celestial labyrinth, each revealing new vistas of insight into the fundamental nature of existence. At the heart of our investigation lies the unveiling of a discrete spectrum of solutions, each showcasing a kaleidoscope of energy states dictating the intricate dance of spin-0 particles within the expansive realm of Bonnor-Melvin's cosmos. Through meticulous quantification of the free equation, buoyed by the introduction of the quantum number $n$, we embark on a journey to decipher the fundamental fabric of reality itself, discerning the myriad pathways through which these particles traverse and interact within the cosmic tapestry.

We aim to investigate the intricate interactions among various objects utilizing the Feshbach-Villars representation, particularly focusing on elucidating their magnetic and thermal characteristics. Breaking away from conventional approaches, we seek to understand the nuanced dynamics governing quantum systems' responses to magnetic fields and temperature changes, with the goal of uncovering underlying mechanisms driving magnetic susceptibility and ordering, as well as gaining fundamental insights into stability and phase transitions. Through thorough theoretical analyses and numerical simulations, we aspire to shed light on novel phenomena and catalyze groundbreaking applications across fields like quantum information science and condensed matter physics. Additionally, we plan to explore oscillators within the Bonnor-Melvin Cosmological model, studying their implications and potential applications, such as the Pöschl-Teller, Morse, Wood-Saxon, q-potential, and pseudoharmonic potentials, utilizing approximation mathematics to deepen our understanding of their mathematical nature and their impacts on particle dynamics and quantum phenomena.

\section*{Funding Statement}

No funds have been received for this manuscript.

\section*{Data Availability Statement}

No new data are generated or analyzed in this study.


\bibliographystyle{model1a-num-names}

\end{document}